\documentclass[preprint,showpacs,amsmath,amssymb,aps,prb,superscriptaddress,lengthcheck]{revtex4-1}

\usepackage{graphicx}
\begin{document}

\title{Wave-diffusion theory of spin transport in metals after ultrashort-pulse excitation}
\author{Steffen Kaltenborn}
\affiliation{Physics Department and Research Center OPTIMAS, University of Kaiserslautern, P.O.Box 3049, 67653 Kaiserslautern, Germany}
\author{ Yao-Hui Zhu}
\affiliation{Physics Department, Beijing Technology and Business University, Beijing 100048, China} 
\author{Hans Christian Schneider}
\email{hcsch@physik.uni-kl.de}
\affiliation{Physics Department and Research Center OPTIMAS, University of Kaiserslautern, P.O.Box 3049, 67653 Kaiserslautern, Germany}


\date{\today}

\begin{abstract}
Spin and charge-current dynamics after ultrafast spin-polarized excitation in a normal metal are studied theoretically using a wave-diffusion theory. It is shown analytically how this macroscopic approach correctly describes the ballistic and diffusive properties of spin and charge transport, but also applies to the intermediate regime between these two limits. Using the wave-diffusion equations we numerically analyze spin and charge dynamics after ultrafast excitation of spin polarized carriers in thin gold films. Assuming slightly spin-dependent momentum relaxation times, we find that a unified treatment of diffusive and ballistic transport yields robust signatures in the spin and charge dynamics, which are in qualitative agreement with recent experimental results [\prl~107, 076601 (2011)]. The influence of boundary effects on the temporal signatures of spin transport is also studied.
\end{abstract}

\maketitle

\section{Introduction}

There has been a persistent interest in spin transport phenomena because of their importance in physics and its application in spintronic devices. Spin transport in magnetic multilayers composed of ferromagnetic metals and simple metals0has been perhaps the most intensely studied of these phenomena:~\cite{zutic,fabian,bauer} for instance, spin injection/detection using currents flowing perpendicular to the plane of the layers (CPP), \cite{silsbee} spin manipulation by electric and magnetic fields~\cite{jedema} and CPP giant magnetoresistance (GMR).~\cite{ValetFert} For a macroscopic description of spin currents and the accompanying spin accumulation at layer boundaries, one often applies macroscopic spin-diffusion theory,~\cite{zhang,rashba,zhang2,sham} which yields an unphysical infinite propagation velocity for a spin signal. This property of the spin diffusion equation becomes troublesome at high frequencies and/or switching speeds, and can be circumvented by using a macroscopic spin \emph{wave-diffusion} equation, as discussed in Refs.~\onlinecite{2008,2009}. In this paper, we are mainly concerned with (i) a general discussion how the wave character of spin transport is related to ballistic transport in the macroscopic spin wave-diffusion theory and (ii) an application of the theory to the case of metallic heterostructures in which the spin polarization is created by ultrashort optical pulses. The motivation for the latter problem comes mainly from recent experimental results,~\cite{melnikov} where spin and charge resolved transport in thin gold films over 100\,nm after ultrashort-pulse excitation was observed and signatures of both ballistic and diffusive transport were found. While the results on charge dynamics in Ref.~\onlinecite{melnikov} bear some resemblance with earlier results,~\cite{Brorson} the measured spin dynamics yield additional characteristic signatures, which we analyze in the framework of spin wave-diffusion equations. Since our theoretical approach contains ballistic and diffusive dynamics as special cases, it is well suited for a study of the behavior at the transition between ballistic and diffusive transport.

The paper is organized as follows. The time-dependent equation system, which forms the basis of our discussion is reviewed in Sec. \ref{sec:TIME-DEPENDENT-EQUATION-SYSTEM}. In Sec. \ref{sec:analytical-solution} we show analytically how the wave character of these equations is related to ballistic transport. In Sec.~IV we present numerical results for the spin and charge dynamics in a gold film after spin-polarized optical excitation, and  compare our results to recent experiments. We summarize our results in Sec.~\ref{sec:CONCLUSIONS}.
\section{Time-dependent equation system\label{sec:TIME-DEPENDENT-EQUATION-SYSTEM}}
We base our discussion of the spin dynamics in a metal on the dynamical equations for the macroscopic spin-current density $J_{s}(z,t)$ and spin density $n_{s}(z,t)$, as derived in Ref.~\onlinecite{2008}: 
\begin{equation}
\label{eq:dnmdt}
\frac{\partial J_{s}(z,t)}{\partial z}+\frac{\partial n_{s}(z,t)}{\partial t}=-\frac{n_{s}(z,t)-n_{-s}(z,t)}{\tau_{\mathrm{sf}}},
\end{equation}
\begin{equation}
\label{eq:dJsdt}
\frac{J_{s}(z,t)}{\tau_{s}}=-c_{s}^{2}\frac{\partial n_{s}(z,t)}{\partial z}-\frac{\partial J_{s}(z,t)}{\partial t}. 
\end{equation}
The equations describe macroscopic spin and charge currents carried by conduction electrons in parabolic bands with spin $s=\pm\frac{1}{2}$. For the description of multilayers it is assumed that the same bands with spin label $s$ exist in each part of the heterostructure. Further, $\tau_{\mathrm{sf}}$ and $\tau_{s}$ are the spin-flip and the (spin-dependent) momentum relaxation times, respectively. Eq.~(\ref{eq:dnmdt}) for the charge current~$J$ and charge density~$n$ has a similar form; only the flip term on the RHS of Eq.~(\ref{eq:dnmdt}) is absent. The equation corresponding to Eq.~(\ref{eq:dnmdt}) involving the charge current and density is just the continuity equation. 

In the derivation of Eq.~(\ref{eq:dnmdt}) from the Boltzmann transport equation, conductivities and momentum relaxation times are introduced to describe dynamical changes of the electronic distributions close to the Fermi energy. In ferromagnets, the momentum relaxation times $\tau_{s}$ are in general different for majority and minority electrons, but in non-magnetic metals they are usually assumed to be spin-independent. To rigorously include excitation further away from the Fermi energy in Eqs.~\eqref{eq:dnmdt} and \eqref{eq:dJsdt}, one would have to augment them by dynamical equations for the electronic energy density (or temperature). We do not do this here, but observe that electrons excited at high energies will relax quickly towards the Fermi energy, so that we can phenomenologically include effects of a \emph{spin-dependent} excitation at energies further away from the Fermi energy by allowing for spin-dependent relaxation times in the non-magnetic metal. Regardless of the excitation conditions, a spin-dependence of the momentum relaxation-times may also arise in an experiment, such as the one in Ref.~\onlinecite{melnikov}, if an external field for the measurement of magneto-optical signals is applied.

To make this paper more self-contained, we collect some results from Ref.~\onlinecite{2008}: There it was shown that an important consequence of the macroscopic wave-diffusion equations is the occurrence of the spin-signal-propagation velocity, 
\begin{equation}
c_{s}=\frac{v_{\mathrm{F}}}{\sqrt{3}},
\label{speed-c}
\end{equation}
where $v_{\mathrm{F}}$ is the Fermi velocity. Eqs.~(\ref{eq:dnmdt}) and (\ref{eq:dJsdt}) can be combined to yield a wave-diffusion equation~\cite{2008} 
for the total spin density $n_{m}=n_{+}-n_{-}$,
\begin{equation}
\frac{\partial^{2}n_{m}}{dt^{2}}+\left(\frac{1}{\tau}+\frac{1}{T_{1}}\right)\frac{\partial n_{m}}{dt}+\frac{n_{m}}{\tau T_{1}}=c_{s}^{2}\frac{\partial^{2}n_{m}}{dz^{2}},\label{eq:spin density}
\end{equation}
where $T_{1}=\frac{\tau_{\mathrm{sf}}}{2}$ is the spin relaxation time and $\tau$ the ``average'' momentum relaxation time. The equation above contains a second-order time derivative, which is absent in the spin diffusion equation.~\cite{fabian} This additional term shows that it takes a finite time for the spin current to adjust to the gradient of the spin accumulation.~\cite{Preziosi,Chester} The second-order time and space derivatives lead to a wave character of the dynamical spin and charge transport in addition to its diffusive character and that time-dependent spin transport should be understood at a macroscopic level in a wave-diffusion duality picture. From this viewpoint, the time-dependent spin diffusion equation can be regarded as an approximation of the wave-diffusion behavior in the long-time/low frequency limit.

\section{Ballistic and diffusive transport from the wave-diffusion equation \label{sec:analytical-solution}}

The wave-diffusion equations described above yield time and space dependent distributions functions, $n_m(z,t)$, that will be discussed in detail in the next section. Here we wish to connect the wave-diffusion picture of spin transport with its diffusive or ballistic character. To this end, we analyze the spatio-temporal behavior of the spin polarized carrier distribution by calculating the mean-square displacement~\cite{Steininger}  
\begin{equation}
\Delta_z^2 (t)  = \frac{1}{N_m} \int_{-\infty}^{\infty}dzz^{2}n_{m}(z,t).
\end{equation}
with $N_m= \int_{-\infty}^{\infty}dz \,n_{m}(z,0)$. We will interpret $\Delta_z^{2}(t)-\Delta_z^2(0) \propto t^2$ as ballistic and $\Delta_z^{2}(t)- \Delta_z^2(0) \propto t$ as diffusive transport in accordance with electronic wave packet dynamics. To simplify the discussion further, we specialize to the case of a spin density that is initially concentrated at $z=0$, i.e., the initial condition 
\begin{equation}
\label{nm0}
n_{m}(z,0)=N_m\delta(z),
\end{equation}
so that $\Delta_z^2(0)=0$.

Following Weiss,~\cite{Weiss} we construct a solution of the wave-diffusion equation with the initial condition~\eqref{nm0} using damped dispersive solutions of the form
\begin{equation}
n_{m}\propto e^{i(kz-\omega t)}.\label{eq:plane wave}
\end{equation}
In Ref.~\onlinecite{2008}, $k(\omega)=k_{r}(\omega)+ik_{i}(\omega)$ was chosen as a complex quantity to obtain the damping length as function of frequency. Here we let $\omega(k)$ become a complex function of the wave vector to determine the temporal damping characteristics. Substituting Eq.~(\ref{eq:plane wave}) into Eq.~(\ref{eq:spin density}), we get the dispersion relation
\begin{equation}
-\omega^{2}-i\alpha\omega+\xi=-c_{s}^{2}k^{2},\label{eq:dispersion relation}
\end{equation}
where $\alpha=\frac{1}{\tau}+\frac{1}{T_{1}}$ and $\xi=\frac{1}{\tau T_{1}}$. We obtain two purely imaginary solutions $\omega_{\pm}=-i\gamma_{\pm}$ with
\begin{alignat}{1}
\gamma_{\pm}\left(k\right)= & \frac{1}{2}\left(\frac{1}{\tau}+\frac{1}{T_{1}}\right)\nonumber \\
&\pm\frac{1}{2}\sqrt{\left(\frac{1}{\tau}+\frac{1}{T_{1}}\right)^{2}-4\left(\frac{1}{\tau T_{1}}+c_{s}^{2}k^{2}\right)},
\end{alignat}
which describe an exponential decay of $n_{m}(z,t)$ and $J_{m}(z,t)$
in time. The initial condition~\eqref{nm0} can now be satisfied by the Fourier integral
\begin{equation}
n_{m}(z,t)=N_m\int_{-\infty}^{\infty}\frac{dk}{2\pi}e^{ikz}\left[e^{-i\omega_{+}(k)t}+e^{-i\omega_{-}(k)t}\right],
\label{eq:plane wave initial}
\end{equation}
from which one reads off the Fourier transformed density
\begin{equation}
\begin{split}
\tilde{n}_{m}(k,t) & = N_m\int_{-\infty}^{\infty}dz e^{-ikz}n_{m}\left(z,t\right) \\
 & =N_m\left(e^{-\gamma_{+}(k)t}+e^{-\gamma_{-}(k)t}\right).
\end{split}
\end{equation}
The mean-square displacement can be expressed with the help of the second derivative of the Fourier transformation and explicitly evaluated as follows 
\begin{equation}
\begin{split}
\Delta_z^{2}(t)  
 =&-\frac{1}{N_m}\left.\frac{\partial^{2}\tilde{n}_{m}(k,t)}{\partial k^{2}}\right|_{k=0}\\
  =&\frac{2c_{s}^{2}t}{\left(\frac{1}{\tau}-\frac{1}{T_{1}}\right)}\left[e^{-t/T_{1}}-e^{-t/\tau}\right], \label{eq:secondmoment}
\end{split}
\end{equation}
where we assumed $\tau <T_{1}$. Eq.~(\ref{eq:secondmoment}) describes the behavior of $\Delta_z^{2}(t)$ for all times.
\begin{figure}
\protect\includegraphics[width=0.40\textwidth]{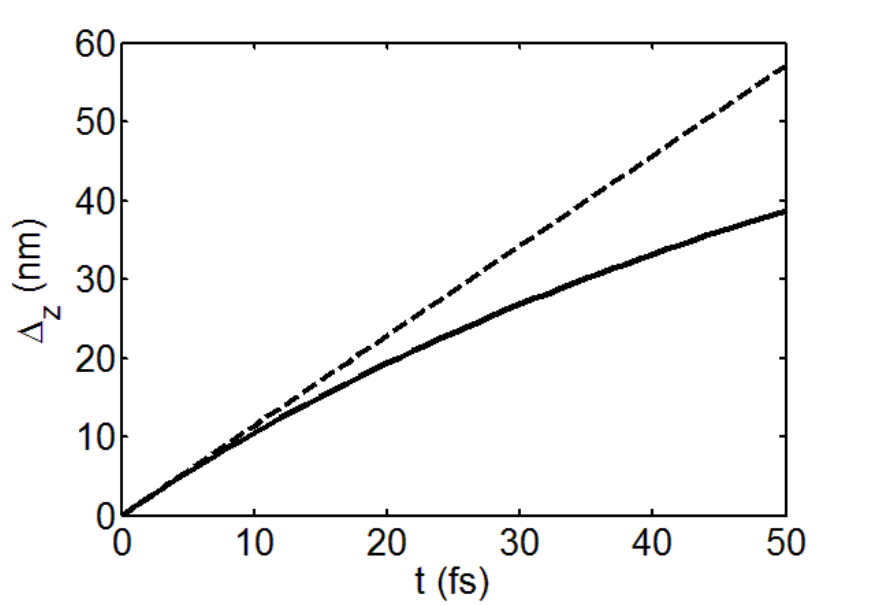}
\caption{\label{fig:variancezoom} Mean-square displacement $\Delta_z(t)$ of the spin density vs. time: Transition from ballistic to diffusive spin transport.}
\end{figure}

\begin{figure}
\protect\includegraphics[width=0.40\textwidth]{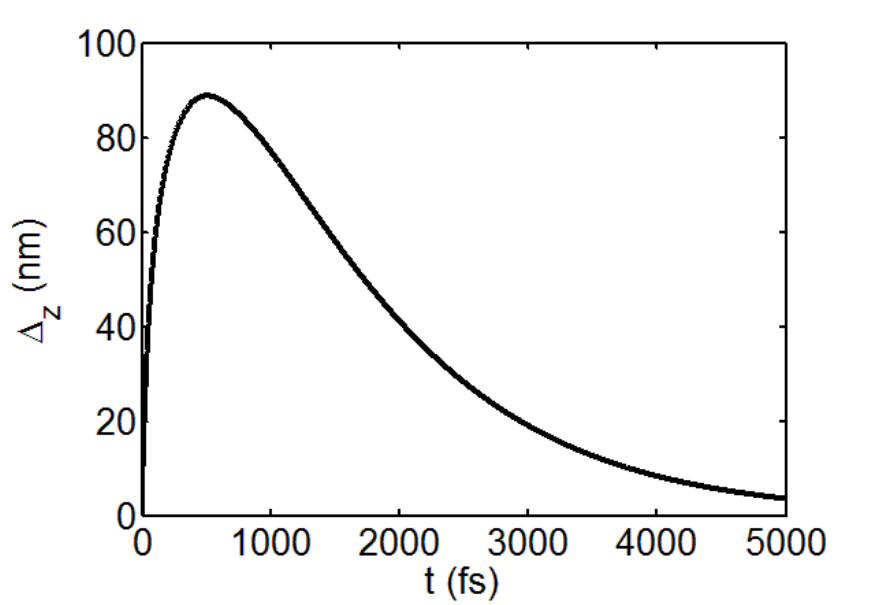}
\caption{\label{fig:variance} Same as Fig.~\protect{\ref{fig:variancezoom}} on a longer time scale.}
\end{figure}

In Fig.~\ref{fig:variancezoom} we plot the time-dependence of the mean-square displacement for the parameters $\left(\tau,T_{1},c_{s}\right)$ of gold (see Sec.~\ref{sec:PARAMETERS-AND-CALCULATIONAL}). On time scales for which the spin relaxation is not important, i.e., $t < T_1$, Eq.~\eqref{eq:secondmoment} simplifies to 
\begin{equation}
\Delta_z^{2}(t)  \simeq 2c_{s}^{2}t\tau\left[1 -e^{-t/\tau}\right].
\end{equation}
For short times, $t \ll \tau$, we have
$\Delta_z^{2}(t)  \propto c_{s}^{2}t^{2}$. The characteristic ballistic behavior of  $\Delta_z^2$ is consistent with the limit $t \ll \tau$, which describes a time scale shorter than typical (momentum) relaxation processes.  For times on the order of the momentum relaxation time $\tau$, momentum scattering processes come into play, and the ballistic behavior is gradually lost. More precisely, for $\tau < t<T_1$, the behavior of the second moment is $\Delta_z^{2} (t)\propto c_{s}^{2}\tau\,t$ with corrections of the order $\exp(-t/\tau)$. By virtue of $c_s^2\tau = D$ and disregarding exponentially small corrections, this means $\Delta_z^2 (t)\propto Dt$, which is a signature of diffusive transport with diffusion constant $D$. Finally, for times on the order of (and longer than) the spin relaxation time~$T_1$ we have an exponential decay of $n_{m}(z,t)$ and also an exponential decay of $\Delta_z^{2}(t)$. This result is simply a consequence of the relaxation of the spin polarization  due to spin-flip processes. These considerations clarify how the macroscopic wave-diffusion equations~\eqref{eq:dnmdt} and \eqref {eq:dJsdt} are capable of describing both the ballistic and diffusive properties of spin transport, as well as the transition between the two limits. We will use these concepts in the discussion of the numerical results for spin transport after ultrashort optical excitation in the following section.

\section{Numerical results}

\subsection{Model and computational setup
\label{sec:PARAMETERS-AND-CALCULATIONAL}}

\begin{figure}
\protect\includegraphics[scale=0.50]{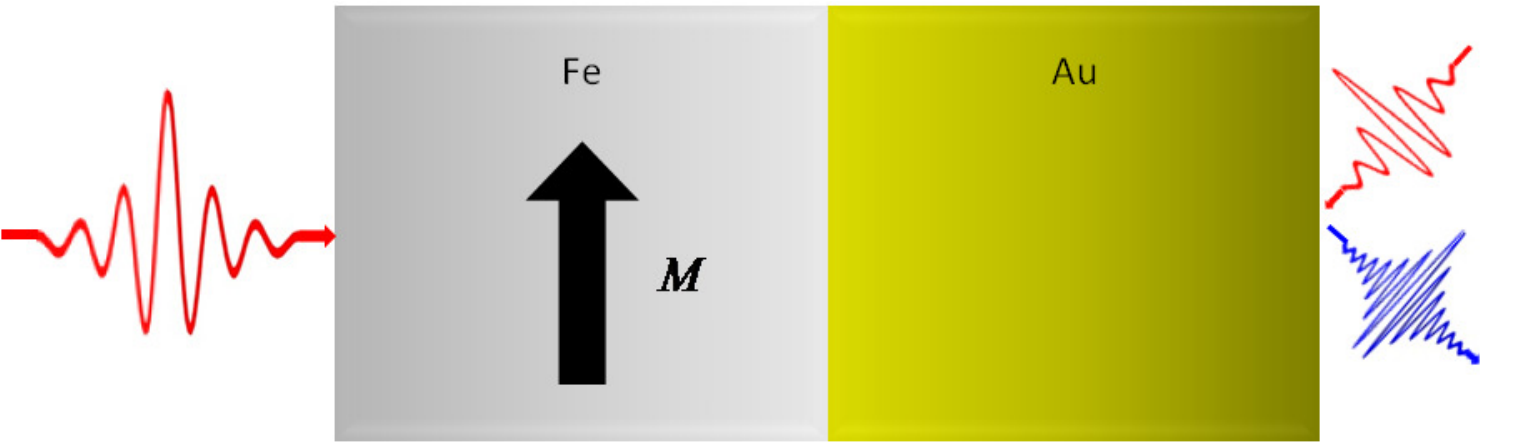}
\caption{ \label{fig:setup} (color online) Schematic setup of the optical pump-probe experiment on a Fe/Au heterostructure.}
\end{figure}

In this section, the time-dependent equation system presented in Sec.~\ref{sec:TIME-DEPENDENT-EQUATION-SYSTEM} is applied to a model setup that aims at analyzing the spin and charge transport as it occurs in the experiment of Melnikov et al.,~\cite{melnikov} as shown in Fig.~\ref{fig:setup}.  
In the experiment, spin polarized carriers are excited optically by a 35\,fs pump pulse in a Fe/Au slab structure. The excited carriers propagate from the iron into the adjacent nonmagnetic gold slab, and the charge and spin dynamics at the right gold surface are extracted from the time-dependent magneto-optical response at the gold surface. 

We are interested in the space- and time-resolved spin and charge dynamics in the gold layer, and introduce the following simplifications.
First, we assume that the spin signal excited in the iron layer can be modeled as a boundary value for the spin and charge densities in the adjacent gold slab. Second, we neglect the spin and charge transport of holes. We then use the macroscopic equations for the electron densities, $n_{s}(z,t)$, and current densities, $J_{s}(z,t)$, in the spin-up and spin-down bands, respectively, of a ferromagnet-metal heterostructure. These dynamical macroscopic equations can be derived from the Boltzmann transport equation for the microscopic carrier distributions under the assumption that the deviations from an equilibrium distribution around the Fermi surface are small.~\cite{2008}  We therefore use the following simplified view of the complicated injection and relaxation process: Spin-polarized carriers are created in the iron layer and move into the spin-degenerate gold bands where this is allowed in the band structure. We assume a very fast initial electronic energy equilibration, so that we have quasi-equilibrium distributions in the spin degenerate bands in gold, but with different densities. The momentum relaxation-times of the quasi-equilibrium electrons of different spin, $\tau_{+}$ and $\tau_{-}$, which arise microscopically from averages over the energy dependence of the electronic distributions, should then also be different. We therefore treat the difference of the relaxation times, $\tau_{+}$ and $\tau_{-}$, as a parameter that needs to be specified. In addition, we could also treat the spin polarization of the injected electrons as a parameter, but we choose not to do so to avoid the introduction of fit parameters. Since it turns out that the dependence of our results on the difference of momentum relaxation-times is extremely weak, the only important parameters of our macroscopic equations have values that are well established from electrical transport measurements: the Fermi velocity $v_{\mathrm{F}}$, the momentum relaxation time $\tau_{s}$ and the spin relaxation time $T_{1}$. Using our approach thus yields the dynamics due to quasi-equilibrium transport, and can be compared to more involved treatments of the relaxation processes away from the Fermi energy. For instance, these relaxation processes may lead to super-diffusive transport.~\cite{oppeneer} 

In order to solve Eqs.~\eqref{eq:dnmdt} and~\eqref{eq:dJsdt} numerically for electronic transport in the gold layer, we need the boundary condition at the interface to the ferromagnet, i.e. the left boundary of our computational domain, which we take to be $z=0$. For simplicity, we assume that the spin-current density as a function of time is known at the Fe/Au interface. This approximation neglects the spin current flowing back to the iron layer due to the spin accumulation in the gold layer. At the Fe/Au interface, $z=0$, the spin-current density is assumed to be a Gaussian
\begin{equation}
J_{s}(z=0,t)=J^{0}e^{-t^{2}/t_0^{2}},
\end{equation}
where $J^{0}$ is the maximum current density. Note that we assume the same initial condition for the currents in the minority and majority channel. Further, $t_0$ is related to the full-width at half maximum of the pulse via $t_0=t_{\mathrm{FWHM}}/(2\log 2)$. We choose $t_{\mathrm{FWHM}}=35$\,fs and $J^{0}=10^{-4}\,\mathrm{nm}^{-2}\mathrm{fs}^{-1}$. For the Au layer, we choose a momentum relaxation time of $\tau_{\mathrm{p}}=30\;\mathrm{fs}$ (from Ref.~\onlinecite{Kruglyak}) and $c_{s}=0.808\;\mathrm{nm}/\mathrm{fs}$ ($v_{\mathrm{F}}=1.4\;\mathrm{nm}/\mathrm{fs}$, see Ref.~\onlinecite{Brorson}). The spin relaxation time  $T_{1}=515\;\mathrm{fs}$ is determined~\cite{ValetFert,2008,difflength} from the spin-diffusion length $l_{\mathrm{sf}}=100$\,nm via the identity $l_{\mathrm{sf}}=c_{s}\sqrt{\tau T_{1}}$. The resulting electron mean-free path is $\lambda=v_{\mathrm{F}}\tau_{\mathrm{p}}=42$\,nm.

In the following, we take the literature value of the momentum relaxation-time for one spin channel, i.e., $\tau_{+}=\tau_{\mathrm{p}}$, but $\tau_{-}\neq\tau_{+}$ to mimic the effect of different microscopic distribution for majority and minority electrons that propagate into the gold layer.  Some justification for our approximation is provided by the weak dependence of the numerical results on the actual value chosen for $\tau_{-}$. 

We study two simplified variations of the setup depicted in Fig.~\ref{fig:setup}. In the first case, we assume an ``infinitely thick'' gold layer to avoid the influence of the reflected spin signal. In this case we assume that the thickness $L$ of the gold slab is so large that spin and charge signals have not propagated to the right boundary during the time interval of interest, i.e., $L\ll l_{\mathrm{sf}}$. In this time frame, the choice of the boundary condition at the right of the gold layer does not influence the results. For mathematical convenience, we require both spin-(charge-) current density and spin (charge) density be zero.
We also study the case of a finite thickness~$L$ of the gold layer. Here, we require the spin and charge current densities to vanish at the right boundary of the gold layer. Although it may seem that this boundary condition applies to the experiment best, it leads to multiple reflections that do not show up in the experiment, as is discussed in more detail below.
 
The numerical solution of Eqs.~(\ref{eq:dnmdt}) and (\ref{eq:dJsdt})  employs the method of characteristics as detailed in Ref.~\onlinecite{2008}.  
 
\subsection{Spin and charge dynamics\label{sec:NUMERICAL-RESULTS}}

\begin{figure}
\includegraphics[width=0.40\textwidth]{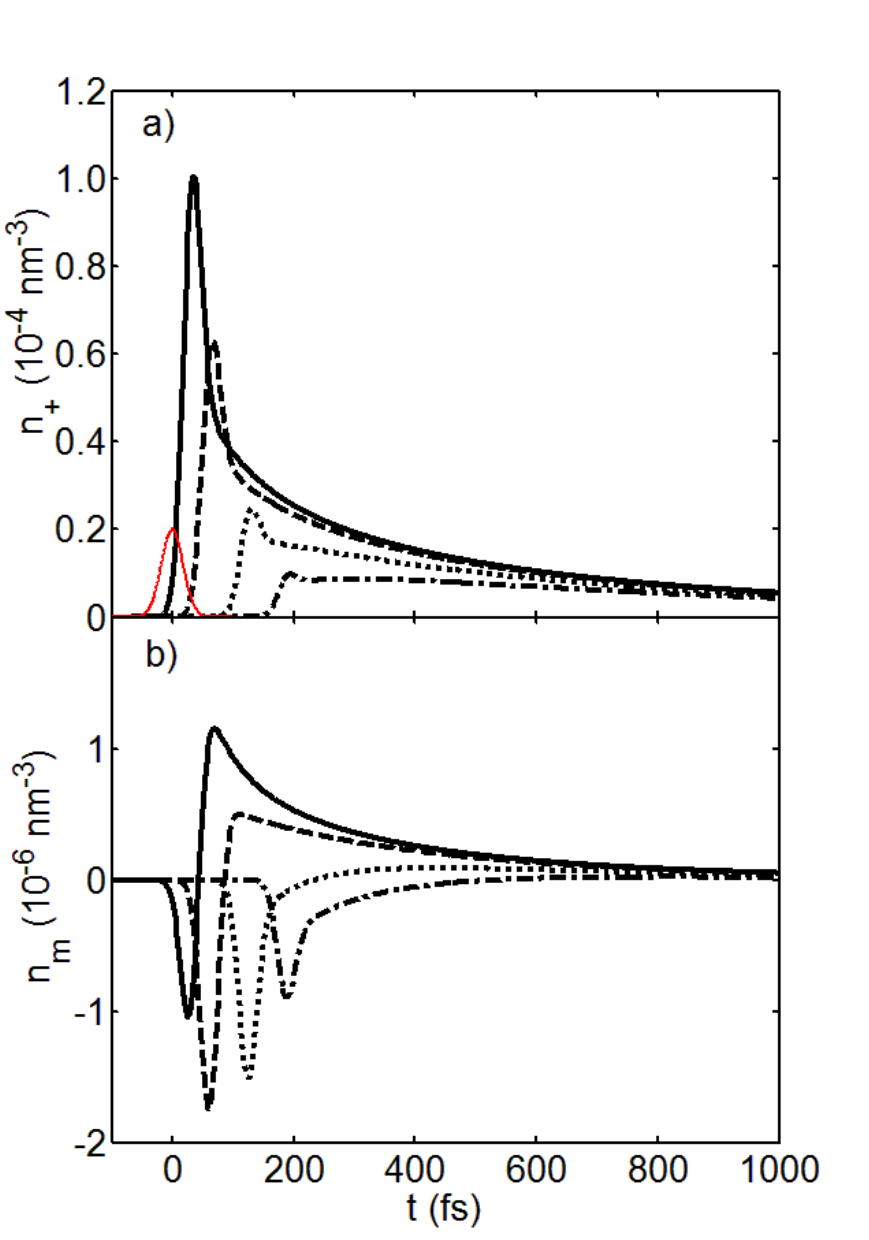}
\caption{\label{fig:infinite-dynamics} Time evolution of the density $n_{+}(z,t)$ of majority electrons (a) and the total spin density $n_{m}(z,t)$ (b) at positions $z=25$\,nm (solid), 50\,nm (dashed), 100\,nm (dotted), and 150\,nm (dash-dotted) of an infinitely thick layer. The excitation current pulse $J_s$ is also shown in arbitrary units(thin red line).}
\end{figure}

We first discuss results for the time evolution of the spin and charge density in an ``infinitely thick'' layer. Fig. \ref{fig:infinite-dynamics}(a) shows the spin density~$n_{+}(z=L,t)$ for majority electrons at different positions $z=L$, ranging from $L=25$\,nm to $L=150$\,nm. As mentioned above, the results at position $z=L$ mimic the dynamics at the surface for a slab thickness~$L$. We find a delayed sharp rise and a peak in the carrier density. We first discuss  the curve for $z=25$\,nm in Fig.~\ref{fig:infinite-dynamics}(a), which is in the regime where ballistic transport properties are dominant, as shown by Fig.~\ref{fig:variancezoom}. The pronounced peak occurs at time $t_{\text{peak}}=L/c=34$\,fs, and therefore is related to wave-like propagation dynamics, which corresponds to ballistic transport as shown in Sec.~\ref{sec:analytical-solution}.  Further, the wave-like dynamics of the carrier distribution preserves to some extent of the Gaussian shape of the driving current pulse $J_s(z=0,t)$, which is created by the drive current at the left boundary. The difference between the full and the purely ballistic result is, of course, due to momentum and spin-flip scattering. These contributions lead to an asymmetric peak and the slow decay, but not to a broadening of the peak. This can be seen by comparing with the result for an infinitely thin layer (not shown), in which the dynamics are completely ballistic.  For peaks at larger distances (and longer times) the peak-like signature is gradually lost as one reaches the regime where the diffusive behavior becomes dominant and the ballistic (wave-like) component is essentially only seen in the delayed onset of the signal.

We therefore interpret the dynamics in Fig.~\ref{fig:infinite-dynamics}(a) as ballistic for short times, as more or less diffusive for ``intermediate'' times, and eventually as an exponential decay for long times. From the ballistic motion of electrons traveling with $v_{\mathrm{F}}$ in a straight line towards the opposite surface one may naively expect the arrival of a signal with the Fermi velocity instead of $c_{s} = v_{\mathrm{F}}/\sqrt{3}$. This is only true for carriers ``going straight'' to the opposite surface, and these carriers are responsible for the onset of the signal. The occurrence of the peak with the smaller velocity of $c$ as well as the finite width of the peak shows that there is a spread in the time the carriers need to traverse a fixed distance in $z$ direction. 

The curve for minority electrons~$n_{-}(z,t)$ (not shown) is almost indistinguishable from the one shown for $n_{+}(z,t)$ on the scale of Fig.~\ref{fig:infinite-dynamics}(a) if we choose the same excitation, i.e., boundary condition, and change the momentum relaxation time for ``$-$'' electrons by 5\% to $\tau_{-}=31.5$\,fs. However, as Fig.~\ref{fig:infinite-dynamics}(b) shows, the total spin density $n_{m}(z,t)=n_{+}(z,t)-n_{-}(z,t)$ is influenced quite drastically by this change. Qualitatively, this can be understood by considering the following picture of the dynamics of individual electrons, even though the calculation deals with distribution functions and the momentum relaxation times are for ensemble averages only: The longer momentum relaxation time results in the minority electrons arriving earlier than the majority ones, which causes a characteristic shape of the spin dynamics with a negative spike in the spin density. When the majority electrons arrive, the signal of the minority carriers already falls off. Therefore the spin density $n_{m}$ changes sign and only afterwards shows a long positive tail. Importantly, the difference between the momentum relaxation times can be changed from 1\% to 100\% leading to changes in the numerical values of the spin density, but not in the qualitative behavior. Thus, as long as there is a difference between $\tau_{+}$ and $\tau_{-}$, the resulting shape of the spin dynamics with the characteristic short negative dip and a long positive decay is extremely robust. 

\begin{figure}
\includegraphics[width=0.40\textwidth]{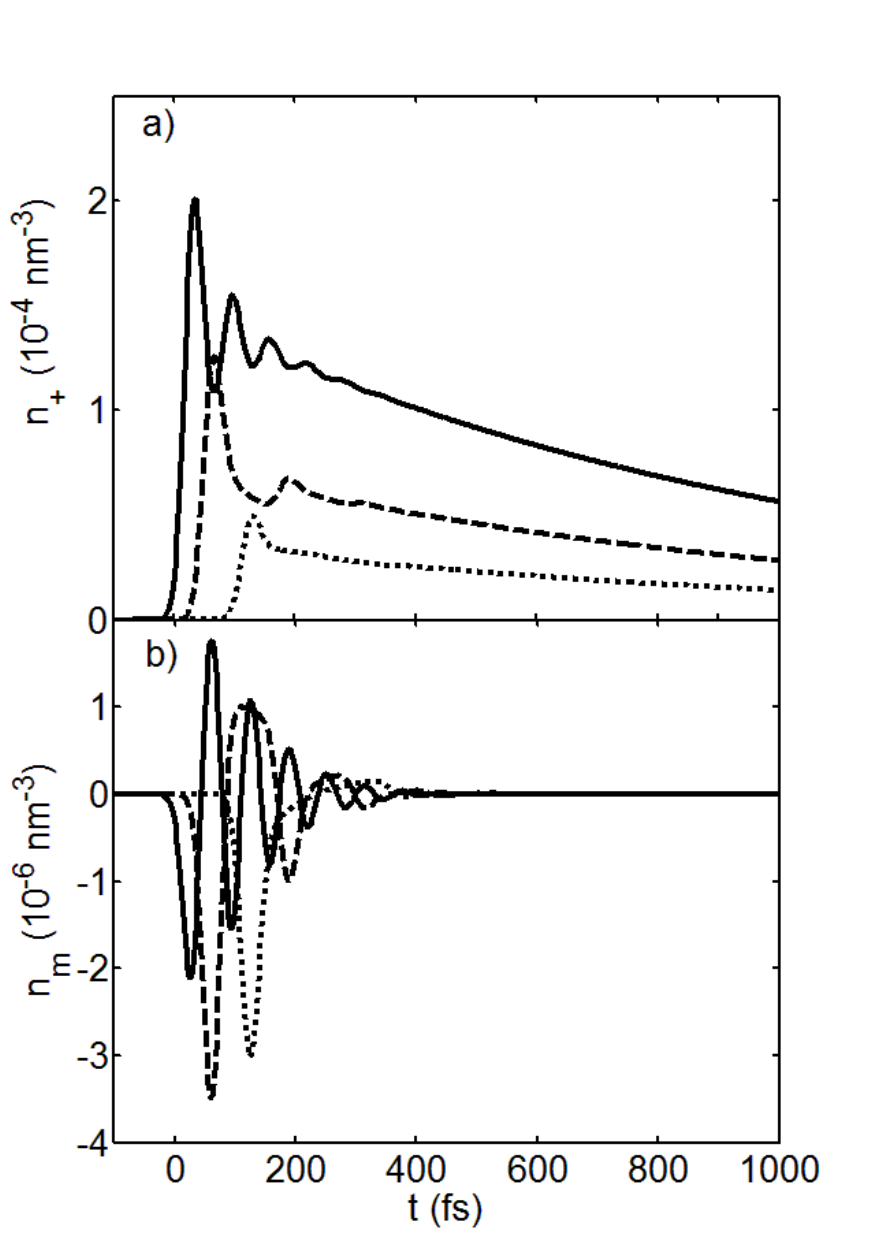}
\caption{\label{fig:time-resolved-spin} Time evolution of the spin density of majority electrons $n_{+}(z=L,t)$ (a) and the total spin density $n_{m}(z=L,t)$ (b) for Au layer thicknesses of  $L=25$\,nm (solid), 50\,nm (dashed), and 100\,nm (dotted).}
\end{figure}

\begin{figure}
\includegraphics[width=0.40\textwidth]{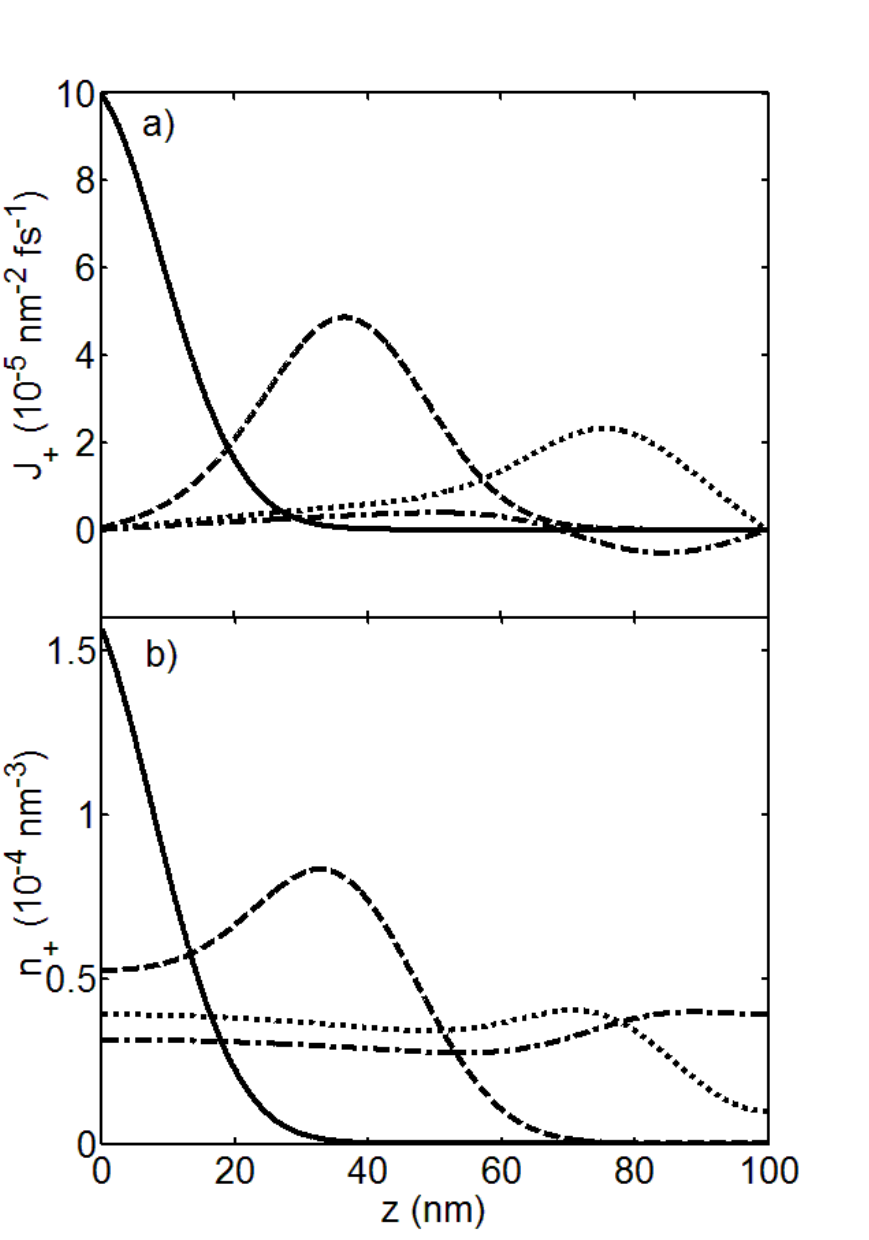}
\caption{\label{fig:space-resolved-spin} Snapshots of the spin-current density $J_{+}(z,t)$ (a) and of the spin density $n_{+}(z,t)$ (b) as a function of $z$ at times $t=0$\,fs, 50\,fs, 100\,fs, and 150\,fs for a $100\,\mathrm{nm}$ gold slab.}
\end{figure}

We now discuss the results for the case in which the finite thickness~$L$ of the metal layer is modeled by a boundary condition for $n_s(z,t)$ at $z=L$. Figure~\ref{fig:time-resolved-spin}(a) shows the spin density $n_{+}(z=L,t)$ for majority electrons at the right boundary of the gold slab as a function of time for slab thicknesses $L=25$, 50, and 100\,nm, respectively.~\cite{note-numerical-values} Several oscillations are visible for the 25\,nm slab. For the 50\,nm slab, only a second peak occurs before the signal decays. This happens because, unlike the case in Fig.~\ref{fig:infinite-dynamics}(a), the electrons are reflected at the right metal surface and continue to propagate/diffuse through the slab. Depending on the thickness of the layer, they may be reflected also at the left boundary, i.e., the Fe/Au interface in the experiment, and propagate again to the right metal boundary. This behavior continues as long as there is a ballistic component to the dynamics and vanishes when the dynamics become diffusive. Eventually, the signal decays on the time scale of the spin relaxation time. In the 100\,nm thick layer, in particular, the ballistic character is further reduced, so that no additional peaks develop at all. The additional peaks, if present in the density dynamics~$n_{+}$ of one spin channel, are also visible in the total spin-density dynamics, $n_{\mathrm{m}}$, as shown in Fig.~\ref{fig:time-resolved-spin}(b). Depending on the thickness of the layer, the spin-signal can change sign more than once for the same reason as without reflection because the propagation and reflection dynamics occurs for both spins with a slightly different relaxation time, so that the maxima and minima of  $n_{-}$ are slightly shifted with respect to those of $n_{+}$. 

Figure~\ref{fig:space-resolved-spin} shows the space dependence of the current and density distributions for ``$+$'' electrons for a slab of thickness $L=100$\,nm at different times. The snapshots of the current density, Fig.~\ref{fig:space-resolved-spin}(a) show how the initial current pulse propagates through the gold slab, is attenuated, and changes sign after it reaches the right slab boundary. The sign change indicates that the propagation direction of the carriers is reversed, i.e., carriers moving ballistically are reflected when they reach the boundary. This is another indicator that even in the case of the 100\,nm gold slab, the ballistic character of the transport is still recognizable at the surface. The spin density $n_{+}(z,t)$, shown in Fig.~\ref{fig:space-resolved-spin}(b), exhibits a wave front at first and a subsequent ``smoothing'' out due to the diffusive character of the dynamics, which becomes dominant after a few hundred femtoseconds. 

\begin{figure}
\includegraphics[width=0.40\textwidth]{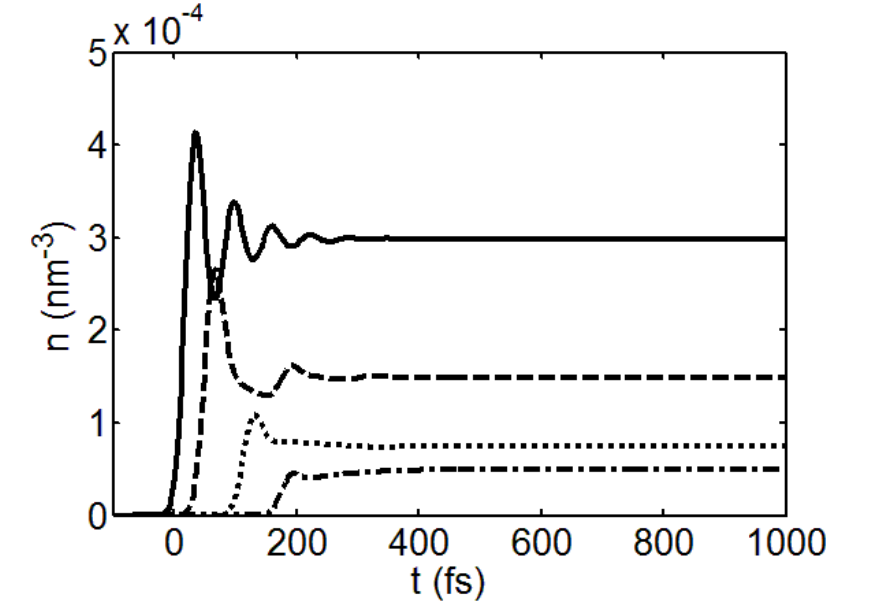}
\caption{\label{fig:charge-dynamics}  Time evolution of the charge density $n(z,t)$ for layer thicknesses of $L=25$\,nm (solid line), 50\,nm (dashed) 100\,nm (dotted), and 150\,nm (dash-dotted).}
\end{figure}

Finally, Fig.~\ref{fig:charge-dynamics} shows the time evolution of the charge density $n(z=L,t)$ at the right boundary of the gold slab for slab thicknesses $L=25$\,nm, 50\,nm, 100\,nm and 150\,nm, respectively. Dependent on the thickness there are additional peaks due to multiple reflections on the boundaries and charge accumulation in the gold layer. The charge accumulates and reaches a steady state because we neglect a possible flowing back of the charge-current into the iron layer.
In an infinitely thick gold layer, the Gaussian current pulse, which mimics the excitation, propagates through the nonmagnetic gold while its maximum decays due to the spin-diffusion length. Thus, the space-resolved results for the spin and charge dynamics show that there is no spin and charge accumulation in an infinitely thick gold layer. 

\subsection{Relation to experiment\label{sec:RELATION-TO-EXP}}

\begin{figure}
\includegraphics[width=0.40\textwidth]{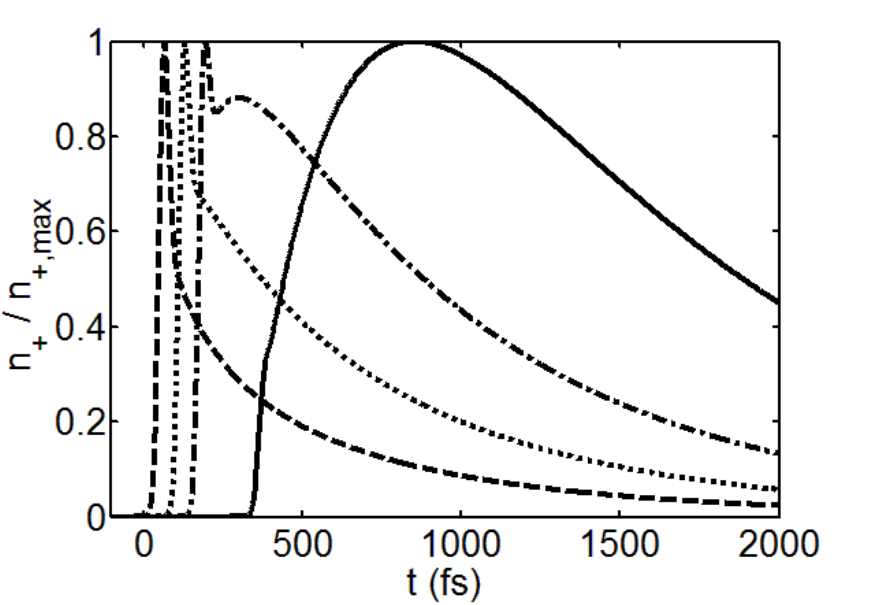}
\caption{\label{fig:scaled-dynamics}  Scaled dynamics of the density $n_{+}(z,t)$ of majority electrons at points $z=50$\,nm (dashed), 100\,nm (dotted), 150\,nm (dash-dotted), and 300\,nm (solid) of an infinitely thick layer.}
\end{figure}

We have already described the idea of the experiment by Melnikov et al., and our calculation applies to the basic setup of that experiment, as shown in Fig. \ref{fig:setup}. It is important to note that we do not calculate the electric fields that are generated by second-harmonic effects at the back surface and whose intensities are measured in the experiment. However, we follow the interpretation of the experimental results by Melnikov et al., and compare the ``charge''  signal with the charge density $n$ and the ``magnetic'' signal with $n_m$. An important experimental finding is a fast rise and a subsequent slow decay of the charge signal accompanied by a magnetic signal that shows a characteristic sign change. Comparing the spin density $n_m$ at the right boundary for slab thicknesses 50\,nm and 100\,nm, see Fig.~\ref{fig:infinite-dynamics}(b) with the \emph{magnetic} SHG signal shown in Figs.~2(c) and (d) of  Ref.~\onlinecite{melnikov} one finds a qualitative agreement. In particular, the time at which the sign change of the signal occurs and the decay at longer times are quite well described by the calculated result. Melnikov et al. observe that their magnetic signal goes to zero at about 1.2\,ps and therefore give roughly 1\,ps as an estimate of the spin-relaxation time (denoted as $\tau_s^{\mathrm{Au}}$ there) pertaining to their experiment. With the help of our model, we can improve this estimate considerably, because our spin relaxation time $T_1 = 515$\,fs is a parameter of the calculation and its numerical value has been determined from transport measurements in the diffusive regime. The calculation therefore provides a link between the ultrafast, optically excited/detected dynamics, and ``conventional'' transport studies. Regarding the degree of agreement, we would like to add two remarks. First, the relative height of the extrema in the calculated spin-density dynamics is somewhat off, but it is possible to improve the agreement with regard to the peak height by making the excitation spin dependent, i.e., by having $J_{+}\neq J_{-}$. This is likely the case in the experiment but including a spin dependent excitation would introduce an adjustable parameter, and we are here not interested in fitting the data, but in analyzing the physics contained in the macroscopic equations~\eqref{eq:dnmdt} and~\eqref{eq:dJsdt}. Second, the calculation seems to lead to sharper extrema and smaller signals at longer times. Very likely, this is due to our neglect of energy relaxation processes because the energy dependence of the distributions underlying the macroscopic equations is assumed to be essentially fixed and concentrated around the Fermi surface.~\cite{2008} The last point is also important when comparing the charge dynamics in our Fig.~\ref{fig:infinite-dynamics}(a)  and Figs.~2(a) and (b) of Melnikov et al. The area under the calculated curves in Fig.~\ref{fig:infinite-dynamics}(a) decreases strongly for larger values of $L$, whereas in the measurement it increases. In the calculation, this is a consequence of the spatial redistribution of the injected carrier density, as can be seen in Fig.~\ref{fig:space-resolved-spin}(b). In the experiment it is probably a result of the energy dependence of the nonlinear process that gives rise to the measured signal, and an energy-redistribution of carriers during the dynamics. It is hard to say which is more important because a microscopic description of the SHG is quite involved.~\cite{Georg} To have a better comparison of the qualitative behavior of charge dynamics for the different slab thicknesses, we normalize the computed charge dynamics in Fig.~\ref{fig:scaled-dynamics}. The normalization not only removes the difference in the peak heights, but also leads to a behavior at long times for both slab thicknesses that is very similar to the experimental result. 

The qualitative agreement of the of the spin and charge dynamics with the signatures seen in the experiment reveals, in our opinion, a view of the  experiment that is complementary to that given by Melnikov et al. There, the energy dependence of the excited carriers was stressed, along with the fact that one needs both electrons and holes with lifetimes that have to match up rather well to be able to separate carriers into those that behave mainly diffusively and those that behave mainly ballistically.  Our simple macroscopic model that contains both a ballistic and diffusive transport properties and using well-established transport parameters suggests that the overall behavior seen in the experiment does not depend critically on details of the electronic lifetimes. Instead, qualitative signatures of the observed dynamics are likely quite generic for transport in thin metal slabs on ultrashort time scales, regardless of the microscopic details of the excitation and relaxation dynamics.

Finally, we compare the general trends seen in Fig.~\ref{fig:scaled-dynamics} with a pioneering measurement~\cite{Brorson} of heat transport in thin gold films excited by a femtosecond laser pulse. This stretches the limits of the applicability of our calculation because in that experiment intraband excitation was likely dominant, leading only to a heating of the conduction electrons and not to an injection of carriers. However, the macroscopic equations for heat transport have the same mathematical properties as the ones used by us,~\cite{Chester,Preziosi} so that one can try to find indications of a qualitative agreement. In this case, we have to compare the normalized dynamics as shown in Fig.~\ref{fig:scaled-dynamics} because their Fig.~2 also shows normalized reflectivity changes. Indeed, the overall shape of the reflectivity changes observed by Brorson et al. match up well with our results. In particular, in Ref.~\onlinecite{Brorson} it is found ``that the rise time of the signal increases slightly with increasing thickness.''  This trend is clearly observed in our calculated results and is explained by a change of dominantly ballistic dynamics in thinner samples to dominantly diffusive dynamics in thicker samples, here 300\,nm. Importantly, this transition comes out of the calculation by a combination of  ballistic and diffusive aspects: Even in thicker samples where the diffusive dynamics dominates the rise time of the signal, there is still an important component of the result that can only be explained by ballistic dynamics, namely the delayed onset of the reflectivity response. Thus it is hard to explain the observed rise times in terms of \emph{either} ballistic \emph{or} diffusive transport, and Brorson et al. noticed this inconsistency in the interpretation of their data. An analysis in terms of wave-diffusion dynamics removes the apparent contradiction between diffusive and ballistic properties of transport.

\section{Conclusions\label{sec:CONCLUSIONS}}

This paper is devoted to an analysis of the wave-diffusion equations, Eqs.~\eqref{eq:dnmdt} and \eqref{eq:dJsdt}, which provide a macrosopic description of carrier transport due to optical or electrical injection. We showed that the wave character in Eqs.~\eqref{eq:dnmdt} and \eqref{eq:dJsdt} is related to ballistic transport by analyzing the behavior of an analytical solution to these equations in one dimension. Scattering processes change the behavior from ballistic at short times to diffusive on a time scale larger than the momentum relaxation time. Importantly, we find that our simple, macroscopic equations work not only in the purely ballistic and diffusive limits, but that they can also applied on time and length scales, where spin and charge transport are neither completely ballistic nor diffusive. Using the wave-diffusion equations, we model the conditions realized in a recent experiment, in which spin-polarized electrons are excited in a gold film by optical pumping of an adjacent ferromagnetic layer and probed optically on the other side of the gold film. If we assume different relaxation times for spin-up and spin-down electrons, our simple macroscopic model can reproduce qualitatively important signatures seen in the experiment, which underscores the importance of the transition between ballistic and diffusive aspects of transport for an understanding of the observed dynamics. Our model uses established macroscopic transport parameters and the dependence of our results on the one additional parameter that we need to introduce, namely the difference of the relaxation times, is very weak. Our results are thus robust, which indicates that the observed behavior is quite generic for ultrafast transport dynamics in thin metal films. Last, we compared the overall behavior of our calculated results to a reflectivity measurement on gold films, and suggest that a unified description of ballistic and diffusive transport can clarify the behavior of electronic transport at ultrashort timescales that may seem unintuitive when one tries to classify transport as either ballistic or diffusive at different times.

\begin{acknowledgments}
We acknowledge support by the state of Rheinland-Pfalz through the MATCOR program and the MAINZ graduate school, as well as by the Beijing Natural Science Foundation (no.~1112007).
\end{acknowledgments}


\begin{thebibliography}{99}
\bibitem{note-numerical-values} The numerical values of $n_s(z,t)$ are twice as large as the ones in Fig.~\ref{fig:infinite-dynamics} because there is no carrier density ``lost'' to propagation outside the slab.
\end{thebibliography}
\end{document}